\documentclass{article}
\usepackage[a4paper]{geometry}
\usepackage[utf8]{inputenc}
\usepackage[T1]{fontenc}
\usepackage[english]{babel}
\usepackage{csquotes}
\usepackage{authblk}
\usepackage{amsmath,amssymb,amsfonts}
\usepackage{amsthm}
\usepackage{braket}
\usepackage{physics}
\usepackage[left]{lineno}
\usepackage{graphicx}
\usepackage{longtable}
\usepackage{tikz}
\usepackage{nicefrac}
\usepackage{quantikz}
\usepackage{algorithm}
\usepackage{algorithmic}
\usepackage{xcolor}
\usepackage{lmodern}
\usepackage{orcidlink}

\usepackage{booktabs}
\usepackage{tabularx}
\usepackage{array}
\usepackage{subcaption}
\usepackage{cleveref}

\crefname{figure}{Figure}{Figures}
\Crefname{figure}{Figure}{Figures}

\title{A Quantum Encoding of Traveling Salesperson Tours via Route Generation, Cost Phases, and a Reversible Valid-Permutation Oracle}

\author[1,2]{Alexander Johannes Stasik\,\orcidlink{0000-0003-1646-2472}\thanks{Corresponding author: \href{mailto:alexander.stasik@sintef.no}{alexander.stasik@sintef.no}}}
\author[1,3]{Franz Georg Fuchs\,\orcidlink{0000-0003-3558-503X}}

\affil[1]{Department for Mathematics and Cybernetics, SINTEF Digital, Oslo, Norway}
\affil[2]{Department of Data Science, Norwegian University of Life Sciences, \AA s, Norway}
\affil[3]{Department of Mathematics, University of Oslo, Oslo, Norway}

\date{\today}

\newtheorem{definition}{Definition}

\newcommand\Order[1]{\ensuremath{\mathcal{O}\left(#1\right)}}

\newcommand\x[0]{\mathbf{x}}

\usepackage{hyperref}

\usepackage[
backend=biber,
style=numeric,
sorting=none,
url=false,
doi=true
]{biblatex}

\begin{document}
\maketitle

\begin{abstract}
For a traveling salesperson problem (TSP) of $n$ cities,
we present a compact quantum encoding based on a time-register representation of tours. A candidate route is represented as a sequence of $n-1$ city labels over discrete time steps, with one fixed start city and the remaining cities encoded in binary registers. We describe three ingredients of the construction: uniform route generation over the route register, a reversible validity oracle, and a phase oracle that encodes the total tour cost. The validity oracle checks both that the non-start city labels form a permutation and, for incomplete graphs, that every directed edge used by the route exists. The cost oracle then accumulates the start-edge, intermediate-transition, and return-edge costs into a tour-dependent phase for valid routes. This yields a coherent superposition of candidate routes with feasibility and tour-length information embedded directly in the quantum state.
The complete construction uses $\mathcal{O}(n\log n)$ qubits, while a naive
implementation requires $\mathcal{O}(n^3\log_2 n)$ CX gates and
$\mathcal{O}\!\left(n^3[\log_2 n+\log_2(1/\epsilon)]\right)$ $T$ gates,
where $\epsilon$ denotes the target approximation precision.
The encoding is compatible with amplitude amplification or spectral filtering techniques such as the quantum singular value transform (QSVT) or Grover's algorithm.
However, due to the exponentially small fraction of valid tours, the overall complexity remains exponential even when combined with amplitude amplification.
\end{abstract}


\section{Introduction}

The Traveling Salesperson Problem (TSP) is a standard benchmark problem in combinatorial optimization \cite{lawler1985traveling}. Given a weighted graph on $n$ cities, the objective is to find a minimum-cost Hamiltonian cycle that visits each city exactly once and returns to the starting city. The number of possible tours grows factorially with $n$.

Quantum algorithms provide a natural framework for representing large combinatorial spaces in superposition. Standard approaches include Grover-style amplitude amplification \cite{grover1996fast,brassard2000quantum}, quantum backtracking and constraint-satisfaction algorithms \cite{montanaro2015quantum,campbell2019applying}, and spectral or polynomial-filtering methods such as quantum singular value transformation (QSVT) \cite{gilyen2019quantum,martyn2021grand}.

Quantum TSP formulations differ in how they represent candidate tours, enforce feasibility, and encode the objective function. Node-time encodings use one-hot variables for city-position assignments and impose the tour constraints through penalties or feasibility-preserving mixers \cite{salehi2022unconstrained,garhofer2024direct}.
Another way is to use direct edge-selection phase encoding in which edge costs are implemented using single-qubit phase gates, while feasibility must be handled separately through state preparation or mixing \cite{garhofer2024direct}.
One can also use a compact binary route Hamiltonian with penalties for repeated cities and unused labels, together with a subspace-reduction encoding based on classical enumeration of feasible tours \cite{madhu2025edge}. These approaches are compared with the time-register oracle construction presented here in \cref{tab:tsp-encoding-comparison}.

\begin{table*}[t]
\centering
\caption{Comparison of representative quantum TSP encodings. The feasible-basis fraction is the fraction of the decision-register basis that represents valid tours for a complete directed graph with one fixed start and end city. It is derived using $N_{\mathrm{tour}}=(n-1)!$ and $b=\lceil\log_2(n-1)\rceil$. Work ancillas are excluded from this fraction.}
\label{tab:tsp-encoding-comparison}
\scriptsize
\renewcommand{\arraystretch}{1.08}
\setlength{\tabcolsep}{3pt}
\begin{tabularx}{\textwidth}{@{}>{\raggedright\arraybackslash}p{0.15\textwidth}>{\raggedright\arraybackslash}p{0.12\textwidth}>{\centering\arraybackslash}p{0.18\textwidth}>{\raggedright\arraybackslash}X>{\raggedright\arraybackslash}p{0.22\textwidth}@{}}
\toprule
Encoding & Qubits & Feasible-basis fraction & Feasibility and cost & Sampling and main drawback \\
\midrule
Node-time one-hot \cite{salehi2022unconstrained,garhofer2024direct} & $(n-1)^2$ & $\displaystyle\frac{N_{\mathrm{tour}}}{2^{(n-1)^2}}=\exp[-\Theta(n^2)]$ & Penalties or feasibility-preserving mixers enforce the constraints; costs couple adjacent positions. & Penalty-based states may contain invalid tours; the encoding also requires quadratic qubit count and $\mathcal{O}(n^3)$ cost interactions. \\
Direct edge-selection phase encoding \cite{garhofer2024direct} & $(n-1)(n-2)$ & $\displaystyle\frac{N_{\mathrm{tour}}}{2^{(n-1)(n-2)}}=\exp[-\Theta(n^2)]$ & Edge costs are encoded using $\mathcal{O}(n^2)$ single-qubit phases; feasibility is handled separately. & Invalid edge sets remain without feasibility-preserving preparation or mixing; no generic mixer construction is provided. \\
Binary route Hamiltonian \cite{madhu2025edge} & $(n-1)b$ & $\displaystyle\frac{N_{\mathrm{tour}}}{2^{(n-1)b}}=\exp[-\Theta(n)]$ & A diagonal many-body Hamiltonian assigns penalties to repeated cities and unused labels. & Invalid strings remain in the basis; the construction requires penalty tuning and high-order interactions. \\
Subspace-reduction encoding \cite{madhu2025edge} & $\lceil\log_2(N_{\mathrm{tour})}\rceil$ & $\displaystyle\frac{N_{\mathrm{tour}}}{2^{\lceil\log_2(N_{\mathrm{tour})}\rceil}}\in(1/2,1]$ & Feasible tours are enumerated classically and inserted into a reduced diagonal Hamiltonian. & The feasible fraction is large, but constructing the encoding requires factorial classical enumeration. \\
Binary time-register oracle, this work & $(n-1)b$ route qubits plus $2n+2$ ancillas & $\displaystyle\frac{N_{\mathrm{tour}}}{2^{(n-1)b}}=\exp[-\Theta(n)]$ & Reversible permutation and edge-validity checks mark feasible routes; cost phases are conditioned on the validity flag. & Uniform preparation includes invalid routes; the dense oracle costs $\mathcal{O}(n^3\log n)$ gates and amplification remains exponential. \\
\bottomrule
\end{tabularx}
\end{table*}

The feasible-basis fraction and the probability of sampling a valid route are distinct quantities. The feasible-basis fraction is a property of the encoding, whereas the sampling probability depends on state preparation, mixing, penalties, postselection, and amplification. A state prepared and evolved entirely within a feasible subspace can produce only valid tours even when the feasible-basis fraction is exponentially small. In contrast, penalty-based formulations generally retain invalid strings in the computational basis.
In the present construction, the validity oracle coherently marks valid routes
by computing the predicate $v(x)$ into an ancillary flag, but does not project
the state onto the valid subspace. Starting from the uniform route state,
invalid routes therefore remain unless the validity flag is subsequently used
for postselection, amplitude amplification, or another filtering procedure.

The construction presented here uses a binary route encoding in the quantum circuit model. Candidate tours are represented as time-ordered sequences of city labels. This representation makes the permutation structure explicit and allows feasibility and tour cost to be implemented through reversible quantum oracles. Unlike penalty-based formulations, unavailable edges and disallowed binary labels are identified directly rather than assigned artificial costs.

We use multi-controlled NOT and multi-controlled single-qubit gates. The multi-controlled NOT gate with $n$ controls is defined as
\begin{equation*}
C^nX
=
\left(
I-\ket{1}\bra{1}^{\otimes n}
\right)\otimes I
+
\ket{1}\bra{1}^{\otimes n}\otimes X.
\end{equation*}
For the decomposition adopted here, $C^nX$ has $\mathcal{O}(n)$ CX count, T-count, and depth and uses one ancilla qubit \cite{Decompositionsofn-qubitToffoliGateswithLinearCircuitComplexity}.

For a control register $\mathbf{c}$, target register $\mathbf{t}$, and control pattern $\mathbf{b}$, a multi-controlled unitary is defined by
\begin{equation*}
C_{\mathbf{b}}^{\mathbf{c}}U_{\mathbf{t}}
=
\left(
I_{\mathbf{c}}
-
\ket{\mathbf{b}}\bra{\mathbf{b}}_{\mathbf{c}}
\right)
\otimes
I_{\mathbf{t}}
+
\ket{\mathbf{b}}\bra{\mathbf{b}}_{\mathbf{c}}
\otimes
U_{\mathbf{t}}.
\end{equation*}
For $U\in SU(2)$ and $n$ control qubits, the decomposition used here has $\mathcal{O}(n)$ CX count and $\mathcal{O}(n+\log_2(1/\epsilon))$ T-count and depth, without ancillary qubits \cite{vale2023decompositionmulticontrolledspecialunitary}. These resource estimates are summarized in \cref{tab:resources}.

\begin{table}[t]
\centering
\begin{tabular}{llll}
\toprule
Gate & \#CX & \#T $=$ depth & \#anc \\
\midrule
$C^nX$ & $\mathcal{O}(n)$ & $\mathcal{O}(n)$ & $1$ \\
$C^nU$, $U\in SU(2)$ & $\mathcal{O}(n)$ & $\mathcal{O}(n+\log_2(1/\epsilon))$ & $0$ \\
\bottomrule
\end{tabular}
\caption{Asymptotic resource requirements for the multi-controlled gates used in this work. For the selected decompositions, T-count and depth have the same asymptotic scaling. The parameter $\epsilon$ denotes the target approximation precision.}
\label{tab:resources}
\end{table}

The contribution of this work is a compact circuit-level baseline for binary route encoding. The construction combines uniform route generation, a reversible validity oracle that checks both permutation validity and directed-edge availability, and a phase oracle that encodes the tour cost only when the validity flag is set. We do not claim a practical quantum speedup. The purpose is to provide an explicit and reproducible encoding for small simulated instances and for possible use with amplitude-amplification or spectral-filtering procedures.

\section{Encoding}

Let $G=(V,E)$ be a weighted directed graph with
\[
V=\{0,1,\dots,n-1\},
\]
and let city $n-1$ be fixed as the start and end city. For every available directed edge $(i,j)\in E$, let $C_{i,j}\geq 0$ denote its traversal cost. A TSP tour is determined by an ordering of the remaining $n-1$ cities together with the requirement that every directed edge used by the ordering belongs to $E$.

We encode the ordering across $T=n-1$ time steps. Each time step stores one non-start city label from
\[
\mathcal{P}=\{0,\dots,n-2\}.
\]
Since city $n-1$ is fixed as the start and end city, it does not need to be encoded in the route register. Let
\[
b=\lceil\log_2(n-1)\rceil
\]
be the number of qubits required to encode one label, and define the set of labels representable by one time register as
\[
\mathcal{D}=\{0,\dots,2^b-1\}.
\]
Only the labels in $\mathcal{P}$ represent non-start cities. If $n-1$ is not a power of two, then $\mathcal{D}\setminus\mathcal{P}$ contains disallowed labels, including the fixed start city $n-1$ and possibly binary values that do not correspond to any city.

The route register is
\[
\mathcal{H}_{\mathrm{route}}
=
\bigotimes_{t=1}^{T}\mathbb{C}^{2^b},
\]
and hence
$
Tb=(n-1)\lceil\log_2(n-1)\rceil
$
qubits are needed for the route encoding. A basis state
\[
\ket{\x}=\ket{x_1}\cdots\ket{x_T}
\]
represents a candidate route. We prepare a uniform superposition over the full route register:
\[
\ket{\psi_{\mathrm{unif}}}
=
\frac{1}{\sqrt{2^{bT}}}
\sum_{\x\in\{0,1\}^{bT}}\ket{\x},
\]
implemented by Hadamard gates on all route qubits. This superposition contains valid tours as well as strings with repeated labels, labels in $\mathcal{D}\setminus\mathcal{P}$, or unavailable graph edges.

\subsection{Validity oracle}

A valid route must satisfy both a permutation condition and an edge-availability condition.

\begin{definition}[Valid route]
A basis state $\ket{x_1,\dots,x_T}$ is valid if
\begin{enumerate}
    \item the multiset $\{x_1,\dots,x_T\}$ equals $\mathcal{P}=\{0,\dots,n-2\}$, and
    \item the directed edges
    \[
    (n-1,x_1),\quad
    (x_t,x_{t+1})\ \text{for }t=1,\dots,T-1,\quad
    (x_T,n-1)
    \]
    all belong to $E$.
\end{enumerate}
\end{definition}

We write
\[
v(\x)=v_{\mathrm{perm}}(\x)\,v_{\mathrm{edge}}(\x),
\]
where $v_{\mathrm{perm}}$ checks the permutation condition and $v_{\mathrm{edge}}$ checks the edge-availability condition.

\subsubsection{Permutation validity}

\begin{figure}[t]
    \centering
    \input{figures/perm_validity_circuit}
    \caption{
        Complete permutation-validity circuit for $n=4$. For each route
        position $t$, the binary values $00$, $01$, and $10$ flip the parity
        ancillas $p_0$, $p_1$, and $p_2$, respectively. The disallowed label
        $11$ produces no parity update. The final three-controlled NOT sets
        $v_{\mathrm{perm}}(\x)=1$ exactly when all three parity ancillas are
        equal to one.
    }
    \label{fig:perm-validity-circuit}
\end{figure}
The permutation-validity predicate is
\[
v_{\mathrm{perm}}(\x)=
\begin{cases}
1, & \text{if }\{x_1,\dots,x_T\}=\mathcal{P},\\
0, & \text{otherwise.}
\end{cases}
\]
For each label $i\in\mathcal{P}$, define its number of occurrences and its parity by
\[
k_i(\x)=\sum_{t=1}^{T}[x_t=i],
\qquad
p_i(\x)=k_i(\x)\pmod 2.
\]

\paragraph{Lemma (automatic exclusion of disallowed labels).}
Because $T=|\mathcal{P}|=n-1$,
\[
p_i(\x)=1\ \text{for every }i\in\mathcal{P}
\]
holds if and only if every label in $\mathcal{P}$ appears exactly once. Consequently, any route containing a repeated label or a label from $\mathcal{D}\setminus\mathcal{P}$ fails the parity test.

\emph{Proof.}
If $p_i(\x)=1$, then $k_i(\x)$ is odd and therefore $k_i(\x)\geq 1$. Hence
\[
\sum_{i\in\mathcal{P}}k_i(\x)\geq |\mathcal{P}|=T.
\]
The left-hand side counts only positions containing labels from $\mathcal{P}$, so it cannot exceed the total number $T$ of route positions. Equality must therefore hold. Thus every $k_i(\x)=1$, and no position contains a label from $\mathcal{D}\setminus\mathcal{P}$. The converse is immediate. \hfill$\square$

Thus, no separate range-check oracle is required: disallowed labels are detected by the permutation-validity test itself.

We implement the test reversibly by introducing one parity ancilla for each non-start city label,
\[
\ket{0}_{\mathrm{par}}^{\otimes(n-1)},
\]
together with a permutation flag qubit $\ket{0}_{\mathrm{perm}}$. For each time step $t$ and each label $i\in\mathcal{P}$, the register $\ket{x_t}$ is compared with the binary encoding of $i$. If they are equal, the parity ancilla associated with $i$ is flipped. A multi-controlled $X$ gate from all parity ancillas onto the permutation flag then gives
\[
v_{\mathrm{perm}}(\x)=1
\iff
p_i(\x)=1\text{ for every }i\in\mathcal{P}.
\]

The equality tests are realized by temporarily bit-flipping controls corresponding to $0$-bits in the binary representation of $i$, applying a $C^bX$ gate to the corresponding parity ancilla, and undoing the temporary $X$ gates. The permutation stage uses $(n-1)^2$ gates of type $C^bX$ and one $C^{n-1}X$ gate. According to \cref{tab:resources}, this requires
$
\Order{n^2\log_2(n)}
$
CX gates and the same asymptotic number of T gates.

For $n=4$, the fixed start city is $3$, there are $T=3$ route positions, and each route label is encoded by two qubits. The labels $00$, $01$, and $10$ represent cities $0$, $1$, and $2$, respectively, while $11$ is disallowed. The circuit in \cref{fig:perm-validity-circuit} computes the parity of each permitted label over all three route positions and then computes the permutation-validity flag. For example, the string $(0,1,3)$ fails because label $2$ occurs zero times, so $p_2(\x)=0$.

\subsubsection{Edge validity}

Let $A$ denote the adjacency matrix of $G$:
\[
A_{i,j}
=
\begin{cases}
1, & (i,j)\in E,\\
0, & (i,j)\notin E.
\end{cases}
\]
For a permutation-valid route, define the $n=T+1$ edge predicates
\[
a_0(\x)=A_{n-1,x_1},
\qquad
a_t(\x)=A_{x_t,x_{t+1}}
\quad (t=1,\dots,T-1),
\qquad
a_T(\x)=A_{x_T,n-1}.
\]
The edge-validity predicate is
\[
v_{\mathrm{edge}}(\x)
=
\bigwedge_{r=0}^{T}a_r(\x).
\]

The computation of each $a_r$ is a reversible implementation of a classical Boolean table lookup. For an intermediate transition, $A_{x_t,x_{t+1}}$ is a Boolean function of the $2b$ bits encoding $(x_t,x_{t+1})$. In direct sum-of-products form,
\[
A_{x_t,x_{t+1}}
=
\bigoplus_{(i,j)\in E_{\mathcal P}}
[x_t=i][x_{t+1}=j],
\qquad
E_{\mathcal P}=E\cap(\mathcal P\times\mathcal P).
\]
The equality indicators in this sum are mutually exclusive: for a fixed basis state, at most one ordered pair $(i,j)$ can match. Consequently, the exclusive OR in the reversible circuit is equivalent to the ordinary Boolean OR of the corresponding minterms. This is the standard sum-of-products representation of switching logic~\cite{Shannon1938}, embedded reversibly as
\[
\ket{i}\ket{j}\ket{z}
\longmapsto
\ket{i}\ket{j}\ket{z\oplus A_{i,j}}.
\]
Reversible embeddings of classical logic are standard~\cite{Bennett1973}, and each minterm can be implemented by a multi-controlled NOT gate using standard decompositions~\cite{Barenco1995}.

A reversible implementation uses one edge ancilla for each of the $n$ tour edges,
\[
\ket{0}_{\mathrm{edge}}^{\otimes n},
\]
and one edge-validity flag $\ket{0}_{\mathrm{edge\mbox{-}valid}}$. The start-edge ancilla is flipped for every available edge $(n-1,i)\in E$, conditioned on $x_1=i$. For each intermediate position $t$, the corresponding edge ancilla is flipped for every $(i,j)\in E_{\mathcal P}$, conditioned on $x_t=i$ and $x_{t+1}=j$. The return-edge ancilla is treated analogously. Because the minterms are mutually exclusive, each edge ancilla is equal to one precisely when the corresponding route edge exists. A $C^nX$ gate from the edge ancillas then computes $v_{\mathrm{edge}}(\x)$.

\Cref{fig:n4-edge-validity} gives a complete example for $n=4$. Open controls denote controls on $\ket{0}$. The first lookup accepts the available start edges $(3,0)$ and $(3,1)$. Each intermediate lookup accepts the three ordered pairs $(0,1)$, $(1,2)$, and $(2,0)$. The return lookup accepts $(0,3)$ and $(2,3)$.

\begin{figure}[t]
    \centering
    \begin{subfigure}[b]{0.38\textwidth}
        \centering
        \begin{tikzpicture}[
            >=stealth,
            city/.style={
                circle,
                draw,
                minimum size=7mm,
                inner sep=1pt
            },
            start/.style={
                city,
                double,
                double distance=1pt
            }
        ]
            \node[start] (n3) at (0,1.8) {$3$};
            \node[city] (n0) at (-1.4,0.2) {$0$};
            \node[city] (n1) at (1.4,0.2) {$1$};
            \node[city] (n2) at (0,-1.4) {$2$};
            \draw[->,bend right=12] (n3) to (n0);
            \draw[->] (n3) to (n1);
            \draw[->] (n0) to (n1);
            \draw[->] (n1) to (n2);
            \draw[->] (n2) to (n0);
            \draw[->,bend right=12] (n0) to (n3);
            \draw[->] (n2) to (n3);
            \node at (0,2.35) {\scriptsize start/end};
        \end{tikzpicture}
        \caption{Directed graph for the $n=4$ example.}
        \label{fig:n4-edge-graph}
    \end{subfigure}
    \hfill
    \begin{subfigure}[b]{0.58\textwidth}
        \centering
        \scriptsize
        \begin{quantikz}[
            row sep=0.20cm,
            column sep=0.20cm
        ]
            \lstick{$x_{1,1}$} & \ctrl[open]{6} & \ctrl[open]{7} & \ctrl[open]{7} & \ctrl{7} & & & & & \\
            \lstick{$x_{1,0}$} & & \ctrl[open]{6} & \ctrl{6} & \ctrl[open]{6} & & & & & \\
            \lstick{$x_{2,1}$} & & \ctrl[open]{5} & \ctrl{5} & \ctrl[open]{5} & \ctrl[open]{6} & \ctrl[open]{6} & \ctrl{6} & & \\
            \lstick{$x_{2,0}$} & & \ctrl{4} & \ctrl[open]{4} & \ctrl[open]{4} & \ctrl[open]{5} & \ctrl{5} & \ctrl[open]{5} & & \\
            \lstick{$x_{3,1}$} & & & & & \ctrl[open]{4} & \ctrl{4} & \ctrl[open]{4} & & \\
            \lstick{$x_{3,0}$} & & & & & \ctrl{3} & \ctrl[open]{3} & \ctrl[open]{3} & \ctrl[open]{4} & \\
            \lstick{$\ket{0}_{a_{\mathrm{start}}}$} & \targ{} & & & & & & & & \ctrl{4} \\
            \lstick{$\ket{0}_{a_{12}}$} & & \targ{} & \targ{} & \targ{} & & & & & \ctrl{3} \\
            \lstick{$\ket{0}_{a_{23}}$} & & & & & \targ{} & \targ{} & \targ{} & & \ctrl{2} \\
            \lstick{$\ket{0}_{a_{\mathrm{return}}}$} & & & & & & & & \targ{} & \ctrl{1} \\
            \lstick{$\ket{0}_{v_{\mathrm{edge}}}$} & & & & & & & & & \targ{}
        \end{quantikz}
        \caption{Edge-validity circuit for the $n=4$ example.}
        \label{fig:n4-edge-circuit}
    \end{subfigure}
    \caption{
    Edge-validity construction for $n=4$, with city $3$ fixed as the
    start and end city. \textbf{(a)} Directed graph used in the example.
    \textbf{(b)} Edge-validity circuit for the same graph. The circuit computes
    $a_{\mathrm{start}}=[(3,x_1)\in E]$,
    $a_{12}=[(x_1,x_2)\in E]$,
    $a_{23}=[(x_2,x_3)\in E]$, and
    $a_{\mathrm{return}}=[(x_3,3)\in E]$.
    The final four-controlled NOT sets
    $v_{\mathrm{edge}}(\x)
    =a_{\mathrm{start}}a_{12}a_{23}a_{\mathrm{return}}$.
    The route $3\to0\to1\to2\to3$ is edge-valid, whereas
    $3\to1\to0\to2\to3$ is marked as edge-invalid because $(1,0)\notin E$.
}
    \label{fig:n4-edge-validity}
\end{figure}

Define
\[
d_{\mathrm{out}}
=
\left|\{j\in\mathcal P:(n-1,j)\in E\}\right|,
\qquad
d_{\mathrm{in}}
=
\left|\{i\in\mathcal P:(i,n-1)\in E\}\right|.
\]
The direct minterm implementation uses
\[
N_{\mathrm{edge}}
=
d_{\mathrm{out}}
+
(n-2)|E_{\mathcal P}|
+
d_{\mathrm{in}}
\]
controlled lookup gates, followed by one $C^nX$ gate that combines the edge ancillas. The start and return lookups have $b$ controls, while each intermediate lookup has $2b$ controls. Under the linear-size decompositions assumed in \cref{tab:resources}, the elementary-gate count is therefore
\[
\Order{
\left(
d_{\mathrm{out}}
+
(n-2)|E_{\mathcal P}|
+
d_{\mathrm{in}}
\right)\log_2(n)
+n
}.
\]
For a dense directed graph, $|E_{\mathcal P}|=\Order{n^2}$, giving the worst-case scaling
$
\Order{n^3\log_2(n)}.
$
For a sparse graph with $|E_{\mathcal P}|=\Order{n}$, the corresponding bound is
$
\Order{n^2\log_2(n)}.
$
These are gate-count estimates; circuit depth additionally depends on the available ancillas, connectivity, and scheduling model. Computing and later uncomputing the edge ancillas changes only the constant prefactor.

The direct minterm network is particularly transparent for small instances. For larger unstructured graphs, the same adjacency predicate may instead be implemented as a coherent table lookup using quantum read-only memory, which provides alternative time-space trade-offs~\cite{Babbush2018}. Such a replacement changes the lookup implementation but not the need to test each of the $n$ edges used by the encoded route.

\subsubsection{Complete validity oracle}

Finally, a Toffoli gate controlled by the permutation-validity and edge-validity flags computes
\[
v(\x)
=
v_{\mathrm{perm}}(\x)\land v_{\mathrm{edge}}(\x)
\]
on a final flag qubit $\ket{0}_{\mathrm{good}}$.

The complete reversible validity computation is
\[
\begin{aligned}
O_{\mathrm{valid}}:\ 
&\ket{\x}
\ket{0}_{\mathrm{par}}^{\otimes(n-1)}
\ket{0}_{\mathrm{perm}}
\ket{0}_{\mathrm{edge}}^{\otimes n}
\ket{0}_{\mathrm{edge\mbox{-}valid}}
\ket{0}_{\mathrm{good}}
\\
\longmapsto\ 
&\ket{\x}
\ket{p(\x)}_{\mathrm{par}}
\ket{v_{\mathrm{perm}}(\x)}_{\mathrm{perm}}
\ket{a(\x)}_{\mathrm{edge}}
\ket{v_{\mathrm{edge}}(\x)}_{\mathrm{edge\mbox{-}valid}}
\ket{v(\x)}_{\mathrm{good}}.
\end{aligned}
\]

For a complete directed graph, $v_{\mathrm{edge}}(\x)=1$ for every permutation-valid route, so the edge-validity stage and its ancillas may be omitted.

The overall implementation uses
$
(n-1)\lceil\log_2(n-1)\rceil+2n+2
$
qubits, consisting of the route register, $n-1$ parity ancillas, one permutation flag, $n$ edge-position ancillas, one edge-validity flag, and one final validity flag. For a complete directed graph, the edge-position ancillas and the edge-validity flag may be omitted.

\subsection{Cost oracle}
For every available directed edge $(i,j)\in E$, let $C_{i,j}\geq 0$ denote its traversal cost. The edge costs need not be symmetric.

\begin{definition}[Tour cost]
For a valid route $\x=(x_1,\dots,x_T)$, the tour cost is
\[
L(\x)
=
C_{n-1,x_1}
+
\sum_{t=1}^{T-1}C_{x_t,x_{t+1}}
+
C_{x_T,n-1}.
\]
\end{definition}

Let $\underline{L}$ and $\overline{L}$ be lower and upper bounds on the cost of every valid route, so that $\underline{L}\leq L(\x)\leq\overline{L}$.
Define the normalized phase angle by
\begin{equation}
\phi(\x)=
\begin{cases}
\displaystyle\frac{L(\x)-\underline{L}}{\overline{L}-\underline{L}}, & v(\x)=1,\\[1ex]
0, & v(\x)=0.
\end{cases}
\end{equation}
The cost oracle applies
\begin{equation}
O_{\mathrm{cost}}\ket{\x}\ket{v(\x)}_{\mathrm{good}}
=
e^{i\phi(\x)}
\ket{\x}\ket{v(\x)}_{\mathrm{good}}.
\end{equation}
Equivalently, because every valid route contains exactly $n$ edge traversals,
the phase-weight matrix may be defined by $\widetilde{C}_{i,j}=(C_{i,j}-\underline{L}/n)/(\overline{L}-\underline{L})$ for $(i,j)\in E$.
The cost oracle then accumulates $\widetilde{C}_{i,j}$ for each traversed
available edge.
Thus, $0\leq\phi(\x)\leq1$ for valid routes, whereas invalid label strings
have $\phi(\x)=0$.
A simple choice is $\underline{L}=n c_{\min}$ and $\overline{L}=n c_{\max}$, where $c_{\min}$ and $c_{\max}$ are the minimum and maximum available edge costs.
Tighter bounds, such as the minimum and maximum directed assignment costs, may also be used.

In circuit form, the oracle applies controlled phase rotations for available edge contributions, with the final validity flag as an additional control. It consists of
\begin{enumerate}
    \item a phase conditioned on $x_1=i$ for every available start edge $(n-1,i)\in E$,
    \item a phase conditioned on $(x_t,x_{t+1})=(i,j)$ for every available intermediate edge $(i,j)\in E_{\mathcal{P}}$ and every $t=1,\dots,T-1$, and
    \item a phase conditioned on $x_T=i$ for every available return edge $(i,n-1)\in E$.
\end{enumerate}
Invalid routes have $v(\x)=0$ and therefore receive no cost phase.

The conditional phase construction is illustrated explicitly in \cref{fig:n4-cost-oracle} for the $n=4$ instance introduced in \cref{fig:n4-edge-validity}.
The figure shows how the start-edge, intermediate-transition, and return-edge cost contributions are implemented as controlled phase rotations on the final validity qubit.
Since the phase gates act on the validity qubit, only routes with $v(\mathbf{x})=1$ accumulate the tour-cost phase, while invalid routes remain unchanged.

\begin{figure}[t]
    \centering
    \scriptsize
    \begin{quantikz}[
        row sep=0.22cm,
        column sep=0.14cm
    ]
        \lstick{$x_{1,1}$}
        & \ctrl[open]{6}
        & \ctrl[open]{6}
        & \ctrl[open]{6}
        & \ctrl[open]{6}
        & \ctrl{6}
        & & & & & \\
        \lstick{$x_{1,0}$}
        & \ctrl[open]{5}
        & \ctrl{5}
        & \ctrl[open]{5}
        & \ctrl{5}
        & \ctrl[open]{5}
        & & & & & \\
        \lstick{$x_{2,1}$}
        & &
        &
        \ctrl[open]{4}
        & \ctrl{4}
        & \ctrl[open]{4}
        & \ctrl[open]{4}
        & \ctrl[open]{4}
        & \ctrl{4}
        & &
        \\
        \lstick{$x_{2,0}$}
        & &
        &
        \ctrl{3}
        & \ctrl[open]{3}
        & \ctrl[open]{3}
        & \ctrl[open]{3}
        & \ctrl{3}
        & \ctrl[open]{3}
        & &
        \\
        \lstick{$x_{3,1}$}
        & &
        &
        &
        &
        &
        \ctrl[open]{2}
        & \ctrl{2}
        & \ctrl[open]{2}
        & \ctrl[open]{2}
        & \ctrl{2}
        \\
        \lstick{$x_{3,0}$}
        & &
        &
        &
        &
        &
        \ctrl{1}
        & \ctrl[open]{1}
        & \ctrl[open]{1}
        & \ctrl[open]{1}
        & \ctrl[open]{1}
        \\
        \lstick{$\ket{v(\mathbf{x})}_{\mathrm{good}}$}
        & \gate{P(\theta_{3,0})}
        & \gate{P(\theta_{3,1})}
        & \gate{P(\theta_{0,1})}
        & \gate{P(\theta_{1,2})}
        & \gate{P(\theta_{2,0})}
        & \gate{P(\theta_{0,1})}
        & \gate{P(\theta_{1,2})}
        & \gate{P(\theta_{2,0})}
        & \gate{P(\theta_{0,3})}
        & \gate{P(\theta_{2,3})}
    \end{quantikz}
    \caption{
        Cost oracle for the $n=4$ graph of
        \cref{fig:n4-edge-validity}. The first two gates encode the
        available start edges, the next three encode the transition
        $(x_1,x_2)$, the following three encode $(x_2,x_3)$, and the final two
        encode the return edge. The validity qubit is the target of each phase
        gate, so invalid routes acquire no cost phase.
    }
    \label{fig:n4-cost-oracle}
\end{figure}

For the $n=4$ graph in \cref{fig:n4-edge-validity}, define
$
\theta_{i,j}=\widetilde{C}_{i,j}, \qquad (i,j)\in E,
$
and let
\[
P(\theta)=
\begin{pmatrix}
1 & 0\\
0 & e^{i\theta}
\end{pmatrix}.
\]
The cost phases can be applied directly to the final validity qubit. Since
$P(\theta)\ket{0}=\ket{0}$, invalid routes acquire no phase. If the validity
qubit is in $\ket{1}$, each control pattern matching an edge used by the route
contributes the corresponding phase.
For example, the valid route
$3\to0\to1\to2\to3$ acquires the phase
\[
\exp\left[
i\left(
\theta_{3,0}
+\theta_{0,1}
+\theta_{1,2}
+\theta_{2,3}
\right)
\right]
=
=e^{i\phi(x)}.
\]

The number of controlled-phase operations is
\[
N_{\mathrm{phase}}
=
d_{\mathrm{out}}
+
(n-2)|E_{\mathcal{P}}|
+
d_{\mathrm{in}}
\leq
2(n-1)+(n-2)(n-1)^2
=
\Order{n^3}.
\]
Using the multi-controlled unitary decompositions summarized in \cref{tab:resources}, the worst-case implementation requires
$
\Order{n^3\log_2(n)}
$
CX gates and
$
\Order{n^3(\log_2(n)+\log_2(1/\epsilon))}
$
T gates. In a combined implementation, the edge-validity and cost phases may reuse the same label and ordered-pair control conditions, reducing constant factors without changing the leading asymptotic scaling.

\section{Overall algorithm}

Let $\ket{0}_{\mathrm{work}}$ denote all parity, permutation, and edge-validity
work qubits. After preparing the uniform superposition over the route register
and applying the validity oracle, the state is
\[
\frac{1}{\sqrt{2^{bT}}}
\sum_{\x\in\{0,1\}^{bT}}
\ket{\x}
\ket{w(\x)}_{\mathrm{work}}
\ket{v(\x)}_{\mathrm{good}}.
\]
The cost oracle then applies the route-dependent phase,
\[
\frac{1}{\sqrt{2^{bT}}}
\sum_{\x\in\{0,1\}^{bT}}
e^{i\phi(x)}
\ket{\x}
\ket{w(\x)}_{\mathrm{work}}
\ket{v(\x)}_{\mathrm{good}}.
\]
Finally, applying $O_{\mathrm{valid}}^\dagger$ uncomputes the validity
workspace while preserving the accumulated phase:
\[
\frac{1}{\sqrt{2^{bT}}}
\sum_{\x\in\{0,1\}^{bT}}
e^{i\phi(x)}
\ket{\x}
\ket{0}_{\mathrm{work}}
\ket{0}_{\mathrm{good}}.
\]

A schematic circuit is
\begin{equation*}
\begin{quantikz}[
row sep={0.55cm,between origins},
column sep=0.65cm
]
\lstick{$\ket{0}_{\mathrm{route}}^{\otimes bT}$}
    & \gate{H^{\otimes bT}}
    & \gate[wires=3]{O_{\mathrm{valid}}}
    & \gate[wires=3]{O_{\mathrm{cost}}}
    & \gate[wires=3]{O_{\mathrm{valid}}^\dagger}
    & \\
\lstick{$\ket{0}_{\mathrm{work}}$}
    &
    &
    &
    &
    & \\
\lstick{$\ket{0}_{\mathrm{good}}$}
    &
    &
    &
    &
    &
\end{quantikz}.
\end{equation*}

The route register consists of $T=n-1$ time registers, each containing
$b=\lceil\log_2(n-1)\rceil$ qubits. The reversible oracle
$O_{\mathrm{valid}}$ computes the permutation-validity and edge-validity
predicates and stores their conjunction in the good flag. The cost oracle
$O_{\mathrm{cost}}$ applies the phase $e^{i\phi(x)}$ when the good flag
is one and acts trivially on invalid routes. The inverse validity oracle then
restores all auxiliary registers to $\ket{0}$, leaving a clean phase encoding
on the route register. This uncomputation is required when the oracle is used
repeatedly inside amplitude amplification or spectral-filtering procedures.

A small numerical example is given in \cref{fig:tsp_example}. The graph in
this example is complete and asymmetric. Therefore, every directed edge is
available, and the edge-validity stage may be omitted.

\begin{figure}[t]
    \centering
    \begin{subfigure}[b]{0.45\textwidth}
        \centering
        \begin{tikzpicture}[
scale=.9,
  >=Stealth,
  node/.style={
    circle,
    draw,
    minimum size=10mm,
    inner sep=0pt,
    font=\sffamily\bfseries,
    fill=gray!15
  },
  edge/.style={
    ->,
    draw=gray!70,
    line width=1.0pt
  },
  weight/.style={
    fill=white,
    inner sep=1pt,
    font=\scriptsize
  }
]

\node[node] (n0) at (18:3.2cm)   {0};
\node[node] (n1) at (-54:3.2cm)  {1};
\node[node] (n2) at (-126:3.2cm) {2};
\node[node] (n3) at (162:3.2cm)  {3};
\node[node, fill=black!50] (n4) at (90:3.2cm)   {4};

\draw[edge]      (n0) to[bend left=16] node[weight, midway, sloped, above] {0.95} (n1);

\draw[edge] (n0) to[bend left=16] node[weight, midway, sloped, above] {0.73} (n2);
\draw[edge] (n2) to[bend left=16] node[weight, midway, sloped, below] {0.02} (n0);

\draw[edge] (n0) to[bend left=16] node[weight, midway, sloped, above] {0.60} (n3);
\draw[edge] (n3) to[bend left=16] node[weight, midway, sloped, below] {0.18} (n0);

\draw[edge]      (n4) to[bend left=16] node[weight, midway, sloped, below] {0.61} (n0);

\draw[edge] (n1) to[bend left=16] node[weight, midway, sloped, above] {0.87} (n2);
\draw[edge] (n2) to[bend left=16] node[weight, midway, sloped, below] {0.97} (n1);

\draw[edge]      (n1) to[bend left=16] node[weight, midway, sloped, above] {0.60} (n3);

\draw[edge] (n1) to[bend left=16] node[weight, midway, sloped, above] {0.71} (n4);
\draw[edge] (n4) to[bend left=16] node[weight, midway, sloped, below] {0.14} (n1);

\draw[edge]      (n3) to[bend left=16] node[weight, midway, sloped, below] {0.53} (n2);

\draw[edge]      (n2) to[bend left=16] node[weight, midway, sloped, above] {0.18} (n4);

\draw[edge] (n3) to[bend left=16] node[weight, midway, sloped, above] {0.29} (n4);
\draw[edge] (n4) to[bend left=16] node[weight, midway, sloped, below] {0.37} (n3);

\draw[edge, red] (n4) to[bend left=16] node[weight, midway, sloped, below] {0.29} (n2);
\draw[edge, red] (n2) to[bend left=16] node[weight, midway, sloped, above] {0.21} (n3);
\draw[edge, red] (n3) to[bend left=16] node[weight, midway, sloped, below] {0.30} (n1);
\draw[edge, red] (n1) to[bend left=16] node[weight, midway, sloped, below] {0.16} (n0);
\draw[edge, red] (n0) to[bend left=16] node[weight, midway, sloped, above] {0.16} (n4);

\end{tikzpicture}
        \caption{Weighted directed graph.}
        \label{fig:tsp-example-graph}
    \end{subfigure}
    \hfill
    \begin{subfigure}[b]{0.50\textwidth}
        \centering
        \begin{tabular}
    {
        >{\raggedright\arraybackslash}p{1.8cm}
        >{\raggedleft\arraybackslash}p{.5cm}
        >{\raggedleft\arraybackslash}p{.3cm}
        >{\raggedright\arraybackslash}p{1.8cm}
        >{\raggedleft\arraybackslash}p{.5cm}
        >{\raggedleft\arraybackslash}p{.3cm}
    }
\toprule
tour & $\phi$ & val & tour & $\phi$ & val \\
\midrule
$[{\color{gray}4,}\, 2, 3, 1, 0{\color{gray},4}]$ & 0.21 & 1 &
$[{\color{gray}4,}\, 1, 3, 2, 0{\color{gray},4}]$ & 0.28 & 1 \\
$[{\color{gray}4,}\, 1, 0, 2, 3{\color{gray},4}]$ & 0.30 & 1 &
$[{\color{gray}4,}\, 1, 2, 3, 0{\color{gray},4}]$ & 0.31 & 1 \\
$[{\color{gray}4,}\, 1, 0, 3, 2{\color{gray},4}]$ & 0.32 & 1 &
$[{\color{gray}4,}\, 3, 1, 2, 0{\color{gray},4}]$ & 0.34 & 1 \\
$[{\color{gray}4,}\, 3, 1, 0, 2{\color{gray},4}]$ & 0.34 & 1 &
$[{\color{gray}4,}\, 1, 3, 0, 2{\color{gray},4}]$ & 0.37 & 1 \\
$[{\color{gray}4,}\, 1, 2, 0, 3{\color{gray},4}]$ & 0.38 & 1 &
$[{\color{gray}4,}\, 2, 0, 3, 1{\color{gray},4}]$ & 0.38 & 1 \\
$[{\color{gray}4,}\, 2, 0, 1, 3{\color{gray},4}]$ & 0.43 & 1 &
$[{\color{gray}4,}\, 3, 2, 1, 0{\color{gray},4}]$ & 0.44 & 1 \\
$[{\color{gray}4,}\, 2, 1, 3, 0{\color{gray},4}]$ & 0.44 & 1 &
$[{\color{gray}4,}\, 2, 1, 0, 3{\color{gray},4}]$ & 0.46 & 1 \\
$[{\color{gray}4,}\, 2, 3, 0, 1{\color{gray},4}]$ & 0.47 & 1 &
$[{\color{gray}4,}\, 3, 0, 1, 2{\color{gray},4}]$ & 0.52 & 1 \\
$[{\color{gray}4,}\, 0, 3, 1, 2{\color{gray},4}]$ & 0.52 & 1 &
$[{\color{gray}4,}\, 0, 2, 3, 1{\color{gray},4}]$ & 0.52 & 1 \\
$[{\color{gray}4,}\, 3, 2, 0, 1{\color{gray},4}]$ & 0.52 & 1 &
$[{\color{gray}4,}\, 0, 1, 3, 2{\color{gray},4}]$ & 0.58 & 1 \\
$[{\color{gray}4,}\, 0, 1, 2, 3{\color{gray},4}]$ & 0.60 & 1 &
$[{\color{gray}4,}\, 3, 0, 2, 1{\color{gray},4}]$ & 0.60 & 1 \\
$[{\color{gray}4,}\, 0, 2, 1, 3{\color{gray},4}]$ & 0.65 & 1 &
$[{\color{gray}4,}\, 0, 0, 0, 0{\color{gray},4}]$ & 0.00 & 0 \\
$[{\color{gray}4,}\, 0, 3, 2, 1{\color{gray},4}]$ & 0.70 & 1 &
\multicolumn{2}{c}{[other invalid tours]} & 0 \\
\bottomrule
\end{tabular}
        \caption{Valid and invalid route strings.}
        \label{fig:tsp-example-table}
    \end{subfigure}
    \caption{
        Illustrative TSP instance with five cities. City $4=n-1$ is the fixed
        start and end city. \subref{fig:tsp-example-graph} Complete directed
        weighted graph defining the cost matrix; the optimal solution is
        shown in red.
        \subref{fig:tsp-example-table} All permutation-valid tours with their normalized phase angles, together with a representative invalid label string.
        In total, the route register contains
        $4^4=256$ encoded candidate strings, of which $4!=24$ represent valid
        tours.
    }
    \label{fig:tsp_example}
\end{figure}

\section{How to use the prepared state}

Let $N_H(G)$ denote the number of directed Hamiltonian cycles in $G$ when city $n-1$ is fixed as the start and end city. Since the route register contains
$
2^{(n-1)\lceil\log_2(n-1)\rceil}
$
basis states, the fraction of valid routes is
\[
p_G
=
\frac{N_H(G)}
{2^{(n-1)\lceil\log_2(n-1)\rceil}}.
\]
For a complete directed graph,
$
N_H(G)=(n-1)!,
$
so
\[
p_G
=
\frac{(n-1)!}
{2^{(n-1)\lceil\log_2(n-1)\rceil}}.
\]
If $n-1$ is a power of two, this becomes
\[
p_G
=
\frac{(n-1)!}{(n-1)^{n-1}}
\sim
\sqrt{2\pi(n-1)}\,e^{-(n-1)}
\]
by Stirling's approximation~\cite{hardy1979introduction}. Consequently, even in the complete-graph case, the number of amplitude-amplification iterations scales as
\[
\Order{\frac{1}{\sqrt{p_G}}}
\sim
\frac{\exp((n-1)/2)}
{(2\pi(n-1))^{1/4}},
\]
and therefore grows exponentially in $n$.

For an incomplete graph,
$
N_H(G)\leq(n-1)!,
$
so the valid fraction is no larger than in the complete-graph case. If $N_H(G)=0$, the graph has no Hamiltonian cycle and the valid subspace is empty. Moreover, $N_H(G)$ is not generally known in advance. Exact phase-adapted amplitude amplification can therefore be used only when the success amplitude is known; otherwise, one must use an amplification procedure that does not require exact prior knowledge of $p_G$.

Once the state has been restricted to the valid subspace, further transformations may favor shorter tours, for example threshold-based amplitude amplification, polynomial filtering, or QSVT-based spectral methods~\cite{brassard2000quantum,gilyen2019quantum}. Missing edges are excluded by the validity oracle rather than represented by large artificial costs. This preserves the phase resolution among admissible tours because the bounds $\underline{L}$ and $\overline{L}$ depend only on the
costs of available edges.

In principle, one could attempt to bias the full superposition toward low-cost assignments before isolating valid tours. However, invalid strings dominate the Hilbert space, and the cost phase is defined to be trivial on the invalid subspace. It is therefore more natural in the present construction to first amplify the valid subspace and then apply cost-selective transformations within it.

\section{Availability of Data and Code}
To support reproducibility and independent verification of our results, we have made all relevant data and source code publicly available at
\url{https://github.com/OpenQuantumComputing/TSP/}.

\section{Funding}
This work was funded by the Research Council of Norway through project number 2656946.

\section{Conclusion}

We presented a compact quantum encoding of TSP tours based on a time-register representation.
The route register encodes candidate orderings directly, the validity oracle
coherently checks and marks both the permutation constraint and the availability
of every directed edge used by the route, and the cost oracle attaches the
objective function as a coherent phase only on the valid subspace.

The displayed construction requires
\[
Q
=
(n-1)\lceil\log_2(n-1)\rceil+2n+2
\]
qubits. These comprise the route register, $n-1$ parity ancillas, one
permutation flag, $n$ edge-position ancillas, one edge-validity flag, and one
final validity flag. Depending on the chosen decomposition of the
multi-controlled gates, an additional reusable work qubit may be required.
For a complete directed graph, the edge-position ancillas and the
edge-validity flag may be omitted.

If
$E_{\mathcal P}=E\cap(\mathcal P\times\mathcal P)$ denotes the set of
available directed edges between non-start cities, then the direct
edge-validity implementation uses
\[
\Order{n|E_{\mathcal P}|\log(n)}
\]
elementary gates, up to lower-order start- and return-edge terms. This gives
\Order{n^3\log(n)} in the dense worst case and \Order{n^2\log (n)} when
$|E_{\mathcal P}|=\Order{n}$. The naive cost oracle has the same dense
worst-case scaling. Thus, adding edge validity for incomplete graphs increases
the constant prefactor but does not change the leading worst-case gate
complexity. For complete graphs, the edge-validity stage is identically
satisfied and may be omitted.

The present construction is primarily representational and does not yield a
polynomial-time quantum algorithm. Even for complete graphs, the fraction of
valid tours is exponentially small, leading to exponential cost under
amplitude amplification. For incomplete graphs, the valid fraction may be
smaller still, or the valid subspace may be empty if no Hamiltonian cycle
exists. Future work should therefore focus on more efficient oracle
constructions, improved state preparation, and alternative encodings that
increase the fraction of valid states. Another direction is the use of
spectral methods such as QSVT to bias the valid subspace toward low-cost tours
without introducing artificial penalties for unavailable edges.

\sloppy
\emergencystretch=1em
\printbibliography

@article{martyn2021grand,
  title = {Grand Unification of Quantum Algorithms},
  author = {Martyn, John M. and Rossi, Zane M. and Tan, Andrew K. and Chuang, Isaac L.},
  journal = {PRX Quantum},
  volume = {2},
  issue = {4},
  pages = {040203},
  numpages = {40},
  year = {2021},
  month = {12},
  publisher = {American Physical Society},
  doi = {10.1103/PRXQuantum.2.040203},
  url = {https://link.aps.org/doi/10.1103/PRXQuantum.2.040203}
}

@article{lawler1985traveling,
  title={The traveling salesman problem: a guided tour of combinatorial optimization},
  author={Lawler, Eugene L},
  journal={Wiley-Interscience Series in Discrete Mathematics},
  year={1985}
}

@inproceedings{grover1996fast,
author = {Grover, Lov K.},
title = {A fast quantum mechanical algorithm for database search},
year = {1996},
isbn = {0897917855},
publisher = {Association for Computing Machinery},
address = {New York, NY, USA},
url = {https://doi.org/10.1145/237814.237866},
doi = {10.1145/237814.237866},
booktitle = {Proceedings of the Twenty-Eighth Annual ACM Symposium on Theory of Computing},
pages = {212–219},
numpages = {8},
location = {Philadelphia, Pennsylvania, USA},
series = {STOC '96}
}

@article{campbell2019applying,
  doi = {10.22331/q-2019-07-18-167},
  url = {https://doi.org/10.22331/q-2019-07-18-167},
  title = {Applying quantum algorithms to constraint satisfaction problems},
  author = {Campbell, Earl and Khurana, Ankur and Montanaro, Ashley},
  journal = {{Quantum}},
  issn = {2521-327X},
  publisher = {{Verein zur F{\"{o}}rderung des Open Access Publizierens in den Quantenwissenschaften}},
  volume = {3},
  pages = {167},
  month = jul,
  year = {2019}
}

@article{montanaro2015quantum,
  title={Quantum walk speedup of backtracking algorithms},
  author={Montanaro, Ashley},
  journal={arXiv preprint arXiv:1509.02374},
  year={2015},
  doi={10.48550/arXiv.1509.02374}
}

@article{brassard2000quantum,
  title={Quantum amplitude amplification and estimation},
  author={Brassard, Gilles and Hoyer, Peter and Mosca, Michele and Tapp, Alain},
  journal={arXiv preprint quant-ph/0005055},
  year={2000},
  doi={10.48550/arXiv.quant-ph/0005055}
}

@inproceedings{gilyen2019quantum,
author = {Gily\'{e}n, Andr\'{a}s and Su, Yuan and Low, Guang Hao and Wiebe, Nathan},
title = {Quantum singular value transformation and beyond: exponential improvements for quantum matrix arithmetics},
year = {2019},
isbn = {9781450367059},
publisher = {Association for Computing Machinery},
address = {New York, NY, USA},
url = {https://doi.org/10.1145/3313276.3316366},
doi = {10.1145/3313276.3316366},
booktitle = {Proceedings of the 51st Annual ACM SIGACT Symposium on Theory of Computing},
pages = {193–204},
numpages = {12},
location = {Phoenix, AZ, USA},
series = {STOC 2019}
}

@book{hardy1979introduction,
  title={An introduction to the theory of numbers},
  author={Hardy, Godfrey Harold and Wright, Edward Maitland},
  year={1979},
  publisher={Oxford university press}
}

@article{vale2023decompositionmulticontrolledspecialunitary,
  title={Decomposition of multi-controlled special unitary single-qubit gates},
  author={Vale, Rafaella and Azevedo, Thiago Melo D and Ara{\'u}jo, Ismael and Araujo, Israel F and da Silva, Adenilton J},
  journal={arXiv},
  year={2023},
  doi={10.48550/arXiv.2302.06377},
}

@article{Decompositionsofn-qubitToffoliGateswithLinearCircuitComplexity,
author = {He, Yong and Luo, Mingxing and Zhang, E. and Wang, Hong-Ke and Wang, Xiao-Feng},
year = {2017},
month = {07},
pages = {},
title = {Decompositions of n-qubit Toffoli Gates with Linear Circuit Complexity},
volume = {56},
journal = {International Journal of Theoretical Physics},
doi = {10.1007/s10773-017-3389-4}
}

@article{madhu2025edge,
  title={An edge-based and subspace reduction encoding scheme to solve the traveling salesman problem in quantum computers},
  author={Madhu, Anandu Kalleri and Li, Chi-Kwong and R{\"o}nkk{\"o}, Jami and Nakahara, Mikio and Lee, Ray-Kuang},
  journal={arXiv preprint arXiv:2512.17291},
  year={2025},
  doi={10.48550/arXiv.2512.17291},
}

@article{garhofer2024direct,
  title={Direct phase encoding in QAOA: Describing combinatorial optimization problems through binary decision variables},
  author={Garhofer, Simon and Bringmann, Oliver},
  journal={arXiv preprint arXiv:2412.07450},
  year={2024},
  doi={10.48550/arXiv.2412.07450}
}

@article{salehi2022unconstrained,
  title={Unconstrained binary models of the travelling salesman problem variants for quantum optimization},
  author={Salehi, {\"O}zlem and Glos, Adam and Miszczak, Jaros{\l}aw Adam},
  journal={Quantum Information Processing},
  volume={21},
  number={2},
  pages={67},
  year={2022},
  publisher={Springer},
  doi={10.1007/s11128-021-03405-5}
}

@article{Shannon1938,
  title = {A symbolic analysis of relay and switching circuits},
  volume = {57},
  ISSN = {2376-7804},
  url = {http://dx.doi.org/10.1109/EE.1938.6431064},
  DOI = {10.1109/ee.1938.6431064},
  number = {12},
  journal = {Electrical Engineering},
  publisher = {Institute of Electrical and Electronics Engineers (IEEE)},
  author = {Shannon,  Claude E.},
  year = {1938},
  month = Dec,
  pages = {713–723}
}

@article{Bennett1973,
  title = {Logical Reversibility of Computation},
  volume = {17},
  ISSN = {0018-8646},
  url = {http://dx.doi.org/10.1147/rd.176.0525},
  DOI = {10.1147/rd.176.0525},
  number = {6},
  journal = {IBM Journal of Research and Development},
  publisher = {IBM},
  author = {Bennett,  C. H.},
  year = {1973},
  month = Nov,
  pages = {525–532}
}

@article{Barenco1995,
  title = {Elementary gates for quantum computation},
  volume = {52},
  ISSN = {1094-1622},
  url = {http://dx.doi.org/10.1103/PhysRevA.52.3457},
  DOI = {10.1103/physreva.52.3457},
  number = {5},
  journal = {Physical Review A},
  publisher = {American Physical Society (APS)},
  author = {Barenco,  Adriano and Bennett,  Charles H. and Cleve,  Richard and DiVincenzo,  David P. and Margolus,  Norman and Shor,  Peter and Sleator,  Tycho and Smolin,  John A. and Weinfurter,  Harald},
  year = {1995},
  month = Nov,
  pages = {3457–3467}
}

@article{Babbush2018,
  title = {Encoding Electronic Spectra in Quantum Circuits with Linear T Complexity},
  volume = {8},
  ISSN = {2160-3308},
  url = {http://dx.doi.org/10.1103/PhysRevX.8.041015},
  DOI = {10.1103/physrevx.8.041015},
  number = {4},
  journal = {Physical Review X},
  publisher = {American Physical Society (APS)},
  author = {Babbush,  Ryan and Gidney,  Craig and Berry,  Dominic W. and Wiebe,  Nathan and McClean,  Jarrod and Paler,  Alexandru and Fowler,  Austin and Neven,  Hartmut},
  year = {2018},
  month = Oct 
}

\end{document}